\def\year{2016}\relax
\documentclass[letterpaper]{article}
\usepackage{aaai16}
\usepackage{times}
\usepackage{helvet}
\usepackage{courier}
\usepackage{multirow}
\usepackage{slashbox}
\usepackage{subfigure}
\usepackage{graphicx}
\usepackage{epstopdf}
\usepackage{times}
\usepackage{epsfig}
\usepackage[latin1]{inputenc}
\usepackage{amssymb}
\usepackage{amsmath}
\usepackage{algorithmic}
\usepackage{algorithm}
\usepackage{verbatim}
\usepackage{multicol}

\newtheorem{Def}{Definition}

\begin{document}
%
\title{Who are Like-minded: Mining User Interest Similarity in Online Social Networks}
\author{Chunfeng Yang\\The Chinese University of Hong Kong\\ Hong Kong, China\\yc012@ie.cuhk.edu.hk 
\And Yipeng Zhou\\Shenzhen University\\Shenzhen, China\\ypzhou@szu.edu.cn
\And Dah Ming Chiu\\The Chinese University of Hong Kong\\ Hong Kong, China\\ dmchiu@ie.cuhk.edu.hk}

\nocopyright

\maketitle
\begin{abstract}
In this paper, we mine and learn to predict how similar a pair of users' interests towards videos are, based on demographic (age, gender and location) and social (friendship, interaction and group membership) information of these users.  We use the video access patterns of active users as ground truth (a form of benchmark). We adopt tag-based user profiling to establish this ground truth, and justify why it is used instead of video-based methods, or many latent topic models such as LDA and Collaborative Filtering approaches. We then show the effectiveness of the different demographic and social features, and their combinations and derivatives, in predicting user interest similarity, based on different machine-learning methods for combining multiple features. We propose a hybrid tree-encoded linear model for combining the features, and show that it out-performs other linear and tree-based models. Our methods can be used to predict user interest similarity when the ground-truth is not available, e.g. for new users, or inactive users whose interests may have changed from old access data, and is useful for video recommendation. Our study is based on a rich dataset from Tencent, a popular service provider of social networks, video services, and various other services in China.

\end{abstract}

\section{Introduction}
\label{Sec:Introduction}
Online social networks (OSNs) have become very popular over the last decade. OSNs connect friends, let them share contents, whether generated by themselves or from public sources, and let them interact. An interesting question is to what extent we can predict the interest similarity of any pair of users, given their demographic and social interaction information available from an OSN provider. The ability for such inference can obviously be very useful, for improving services and making various recommendations to users ~\cite{lewis2012social,yu2014facebook}. In this paper, we focus on the inference of users' interest similarity for videos, based on various information gathered from OSNs.

Typically, a video service provider may only have users' past video consumption data, and use that for developing users' video interest profiles. There are various techniques, notably collaborative filtering (CF) and its variants~\cite{su2009survey}. In practice, the adoption of CF is not without its challenges, especially in a system continuously with a large percentage of new and inactive users, and high rate of content churn and change in user interests. Our study of inferring user interest similarity based on other user information (apart from video consumption data), if the results are positive, can greatly help remedy the situation.

Without content categorization (which can be done by CF or other methods), computing and comparing two users' interest similarity directly from the items (videos) they consumed in the past can be quite misleading. For example, two users may have similar interests, but due to the large stock of videos available and the users only consumed a tiny sample from the entire population, there is no overlap in the videos they actually consumed. This leads to the wrong conclusion that the users have no common interests. So a prerequisite to any study of user interest similarity is a system to categorize the items (videos). In our study, we rely on a tag-based video categorization system, because it is already available from the data source. Each video can have multiple tags. The set of tags are manually generated by multiple editors, which can be considered as a folksonomy approach. We develop two methods to count the tagging information for the purpose of comparing user interests. One method is based on the frequency of common tags between users, called Popular Tag based Profiling (PTP); and the other method takes into account the representativeness of the tags, called Representative Tag based Profiling (RTP).

Given PTP and RTP, we can systematically study the interest correlation of active users (because we have their video consumption data) for different features, including demographic (e.g. age, gender, location) or social ones (friendship, community membership, and interaction information). We also studied the similarity of a user's interest with her past; and with the average of all users. This is the first part of our results. The results show that we can effectively identify the usefulness of different features, and some of them can help infer user similarity effectively.

Then we apply different machine learning models, including linear models, tree-based models, and a hybrid model developed by ourselves, to learn how to best use the available features to infer user interest similarity when we have no consumption data. This is the second part of our results. The results show that our hybrid method out-performs the others. 

To demonstrate the benefit of our results in real applications, we further apply the similarity inferred from our models to video recommendation, and implement several benchmarks for comparison. The experiment results validate the strength of our prediction model and the tag-based user profiling scheme in real applications. Moreover, While PTP can result in more accurate recommendations, RTP can produce more diverse recommendations.

Our study is based on data from our collaborator, Tencent, a large OSN provider that also provides other online services, including video streaming.

The rest of this paper is organized as follows. Sec~\ref{Sec:data} describes the system and data we studied. We define the two tag-based user profiling methods, PTP and RTP, and the corresponding interest similarity calculation in Sec~\ref{Sec:tagProfiling}. We then investigate the correlation between various user information and interest similarity in Sec.~\ref{Sec:measurement}. Experiments of interest similarity prediction are detailed in Sec.~\ref{Sec:simPrediction}, while Sec.~\ref{Sec:recommendation} further applies the predicted results into video recommendation. We conclude in Sec.~\ref{Sec:conclusion}, and related works are summarized and discussed in Sec.~\ref{Sec:related}.

\section{System Overview and Data Description}
\label{Sec:data}
We briefly introduce the system we studied and the available data in this section.

Our collaborator, Tencent Inc., runs an online platform that provides multiple services, including online games, online videos, and instant messaging (QQ), and facilitates the formation of QQ groups. Tencent QQ is the most popular Instant Messaging service in China with roughly 843 million active QQ accounts in June 2015. Users in Tencent QQ can make friends, chat with friends and join QQ groups. QQ groups allow users to easily communicate within a small circle of typically 50 to 100 users, sharing common interests. Tencent Video is one of the largest online video providers in China, supporting millions of daily active users. Tencent Video's video catalog includes movies, TV episodes, music videos, news, user generated content, and more. 
In this paper, we analyzed comprehensive user-related information: 
\begin{itemize}
\item Demographic information: Demographic information includes age, gender, location, occupation and income. On analyzing the most users, we used three types of demographic information, namely, gender, age, and location.
\item Social relationship: In this paper, we considered friendship, social interaction and group membership of users.
\item Interest data: This work focused on user interest for videos. We utilized user's video access records over time.
\end{itemize}
To help discuss our ideas precisely, we define some notations.
The sets of users and videos are denoted by $\mathcal{U}$ and $\mathcal{I}$, respectively. Total number of users is $\left|\mathcal{U}\right|$. We use $u$ and $v$ as the indices of users, and $m$ and $n$ as the indices of videos. The gender, age and location of a user $u$  are represented by $g_u$, $a_u$ and $l_u$. User $u$'s friend set is $\mathcal{F}_u$ and we denote the friendship between user $u$ and $v$ as $F(u,v)$, where $F(u,v)=1$ if $u \in \mathcal{F}_v$  and $v \in \mathcal{F}_u$. Note, the friendship in Tencent QQ is reciprocal. The QQ groups joined by user $u$ is denoted as $\mathcal{G}_u$. Interaction (messaging) between friends on a certain day is represented by $m_d(u,v)$, where $d$ is the index of the day (if we index the current day as 0, then the past day is -1, and the like). The viewed video set of user $u$ is $\mathcal{I}_u$\footnote{Notations related to users' viewing behaviors all refer to one day's data unless  stated otherwise,  and for simplicity, the time stamp is omitted if there is no ambiguity.}, which in fact is the traditional video-based profile of user $u$.

\section{User Interest Profiling based on Tags}
\label{Sec:tagProfiling}
In traditional CF with implicit feedback data, similarity between users is defined as the ratio of common video consumption. However, the available set of videos is so large in most online video systems that two users may share no common video access even if they have similar interests. Thus, we proposed a tag-based profiling scheme which is more general than the video-based profiling, that is, two users who are similar under video-based profiling also have a large similarity value using the tag-based profiling, and the reverse is not necessarily true.  
To achieve this goal, Tencent Video employed many users (``editors'') dedicated to viewing and labeling videos with one or more tags from a predefined tag vocabulary (in order to avoid colloquial and noisy tag usage). For each video, tags used by more than a certain proportion of editors are kept.
Currently, there are around 30 thousand tags in the tag vocabulary, and some examples are ``visually intriguing'', ``reality show'' and ``mixed feelings''. We then generate tag-based user profiles by fusing the user-video consumption data and video-tag relations, that is, converting the user-video-tag tripartite graph into a user-tag bipartite graph. In this way, users consuming the same video will be labeled with the same set of tags. Note that another potential approach to alleviate the weakness of video-based profiling is to describe users' interest by some latent topics derived by matrix factorization~\cite{koren2009matrix} or LDA~\cite{krestel2009latent} from the video consumption and description data. However, the latent topic based approaches require large-scale computation efforts, which needs to be repeated as new user-item consumptions come. Thereby, it's more efficient to utilize tag-based methods in practical systems, such as, Tencent Video.

\begin{figure}[htb]
  \centering
  \subfigure[]{
    \label{Fig:user_tag_count}
     \includegraphics[width=1.6in]{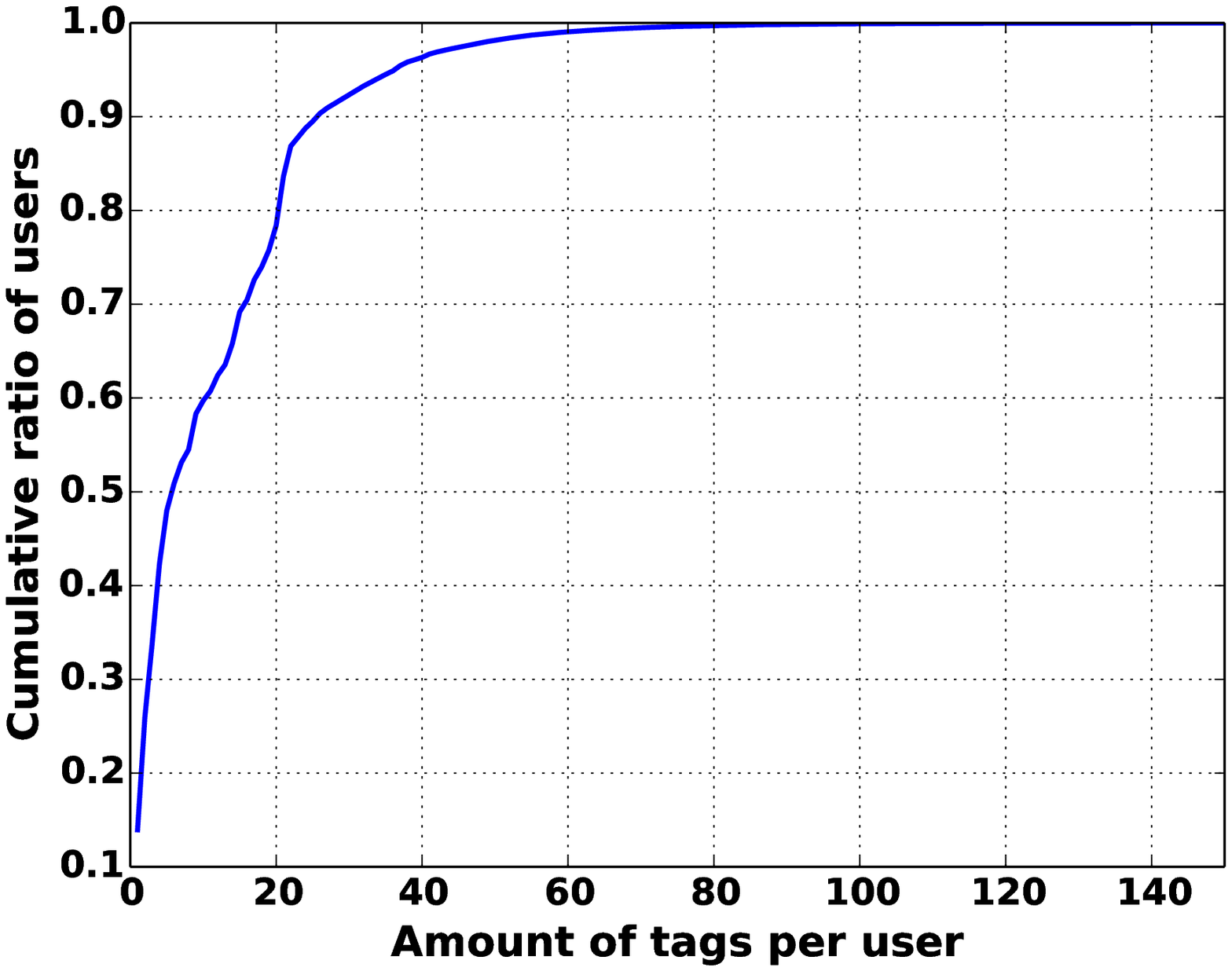}}
  \subfigure[]{
    \label{Fig:tag_popularity} 
		 \includegraphics[width=1.6in]{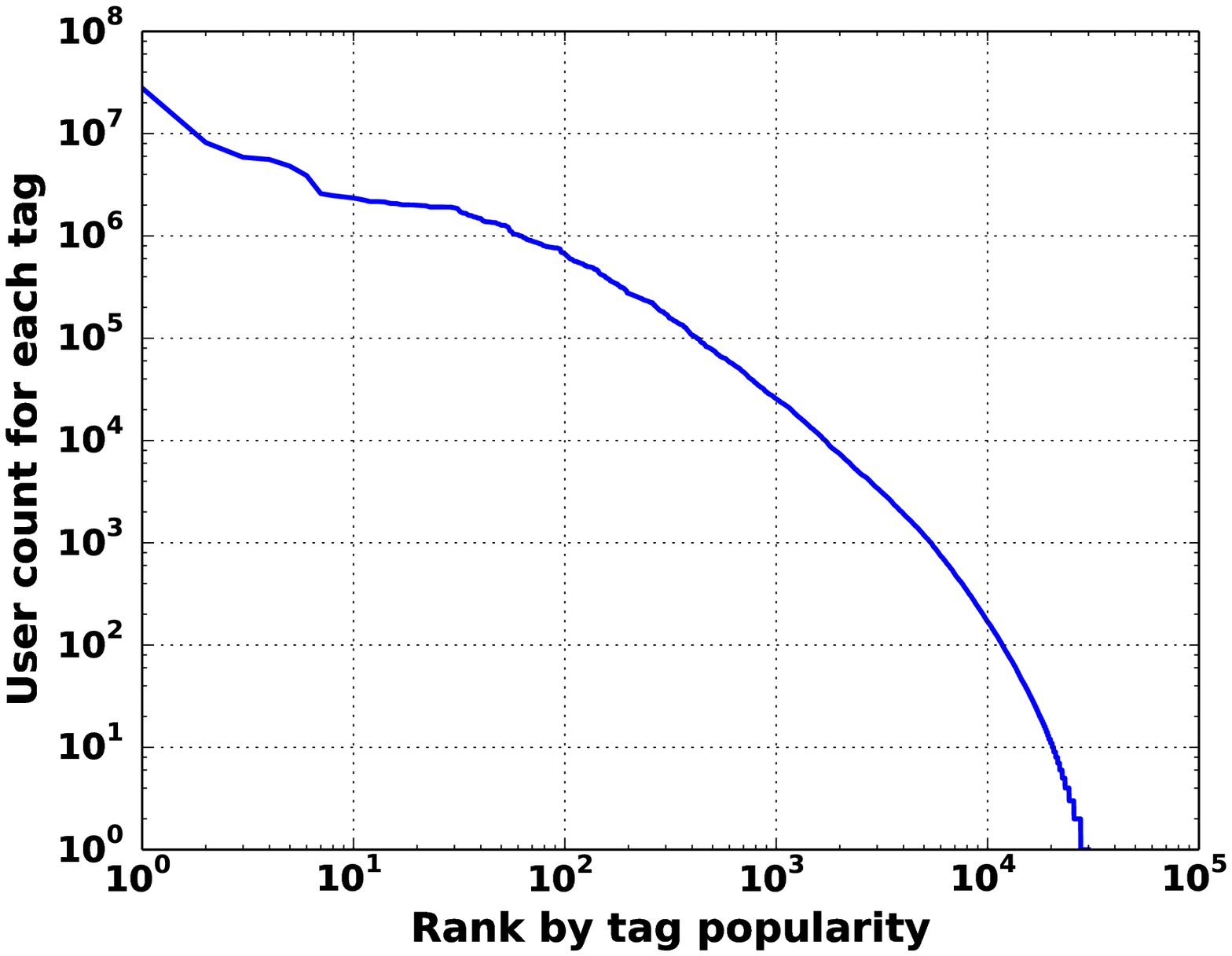}}
	\caption{(a). Distribution of tag count per user (one day); (b). Rank of user count for each tag (tag popularity) per day.}
  \label{Fig:data} 
\end{figure}

In this paper, we are interested in users' current interest similarity that is more accurately captured by recent viewing behaviors, as justified by the observation of interest evolution in the Sec.~\ref{Sec:measurement}. We define active users as ones who have viewed one or more videos in a recent target duration, such as a week or a day, and inactive users as those without viewing records during this period. Note that it may be more precise to call them inactive users rather than new users because they may be active in the past. And it's common in Tencent Video that users who were once active become inactive during a certain recent period. Without loss of generality, we use daily data to calculate the current interest similarity. 
We further define some tag-related notations.
The set of tags is represented by $\mathcal{T}$. We use $i$ and $j$ as the indices of tags. The tag set of video $m$ is denoted as $\mathcal{T}_m$. The tag set of user $u$ is $\mathcal{T}_u=\{i|i \in \mathcal{T}_m, m \in \mathcal{I}_u \}$, and the set of users who have tag $i$ is $\mathcal{U}_i=\{u|i \in \mathcal{T}_u\}$. The distribution of the sizes for users' tag sets, i.e., $|\mathcal{T}_u|$, is in Fig.~\ref{Fig:user_tag_count}.
\begin{Def}
(Popular Tag based Profiling, PTP). The weight of tag $i$ is proportional to the number of videos labeled by tag $i$ and viewed by user $u$ during a certain period. The user profile, obtained by aggregating the tag sets of videos viewed by $u$, is denoted by\footnote{Since each tag in the profile has a weight, we use a dictionary-like structure to represent the profile, that is, $w_i^P = \mathbb{T}_u^P\left[i\right]$. The same structure is adopted for profiles in RTP.}
\begin{eqnarray}
\begin{aligned}
\mathbb{T}_u^P = \left\{\left( i:w_i^P\right)|w_i^P = \bigl|\left\{ m | i \in \mathcal{T}_m, m \in \mathcal{I}_u\right\}\bigr|  \right\}
\end{aligned}
\end{eqnarray}
\end{Def}

The distribution of the number of users owning tag $i$, i.e., $\left|\mathcal{U}_i\right|$, is shown in Fig.~\ref{Fig:tag_popularity}. We can observe that the distribution is quite skewed indicating that there are some popular tags held by numerous users. Similar to TF-IDF in information retrieval, to identify more informative and representative tags, we propose another tag-based profiling method by penalizing tags that are very popular among user profiles. The RTP method is effective in exploring users' personal long-tail interests, which will be shown in the experiments.

\begin{Def}
(Representative Tag based Profiling, RTP). Besides user $u$'s individual preference of tag $i$, RTP also considers the occurrence of tag $i$ in all users' tag lists. The user profile is represented as
\begin{eqnarray}
\begin{aligned}
\mathbb{T}_u^R = \left\{ \left( i:w_i^R \right) | w_i^R = w_i^P*log_2 \frac{|\mathcal{U}|}{|\mathcal{U}_i|}, w_i^P = \mathbb{T}_u^P\left[i\right] \right\}
\end{aligned}
\end{eqnarray}
\end{Def}

For each user profiling method, namely, PTP, RTP and video-based profiling, we use cosine similarity measure to calculate the similarity:
\begin{eqnarray}
\label{eq:PTP_sim}
\begin{aligned}
S^P\left(u,v\right)=\frac{\sum_{i\in \mathcal{T}_u \cap \mathcal{T}_v}\mathbb{T}_u^P[i]*\mathbb{T}_v^P[i]}{\sqrt{\sum_{i\in \mathcal{T}_u}(\mathbb{T}_u^P[i])^2}*\sqrt{\sum_{i\in \mathcal{T}_v}(\mathbb{T}_v^P[i])^2}}
\end{aligned}
\end{eqnarray}
\begin{eqnarray}
\label{eq:RTP_sim}
\begin{aligned}
S^R\left(u,v\right)=\frac{\sum_{i\in \mathcal{T}_u \cap \mathcal{T}_v}\mathbb{T}_u^R[i]*\mathbb{T}_v^R[i]}{\sqrt{\sum_{i\in \mathcal{T}_u}(\mathbb{T}_u^R[i])^2}*\sqrt{\sum_{i\in \mathcal{T}_v}(\mathbb{T}_v^R[i])^2}}
\end{aligned}
\end{eqnarray}
\begin{eqnarray}
\label{eq:vid_sim}
\begin{aligned}
S^I\left(u,v\right)=\frac{\left|\mathcal{I}_u\cap \mathcal{I}_v\right|}{\sqrt{\left|\mathcal{I}_u\right|}*\sqrt{\left|\mathcal{I}_v\right|}}
\end{aligned}
\end{eqnarray}

Note that, we are not the first one to propose tag-based user profiling, however, the novelty lies in conducting extensive empirical studies, similarity prediction and video recommendation experiments under the two different profiling methods as shown in the following sections.

\section{Correlation Study}
\label{Sec:measurement}
In this section, we explore correlations between various user features and interest similarity, and try to seek key features that influence interest similarity through extensive empirical study.
\subsection{Correlation of User Demographics to Interest}
We analyze how demographic information affects interest similarity in terms of PTP and RTP. We consider three demographic features: gender, age and location.
\subsubsection{Gender.} 
We measured the average interest similarity between user pairs with the same or different genders, and the results are shown in Table~\ref{tab:gender}.

\begin{table}[htb]
\centering
\caption{Average interest similarity vs. gender of two users}\label{tab:gender}
\begin{tabular}{|c|c|c|}
\hline
 & PTP  & RTP \tabularnewline
\hline
Male-male & 0.1165  &   0.0562 \tabularnewline
\hline
Male-female & 0.1305 &  0.0698 \tabularnewline
\hline
Female-female & 0.1643  &  0.0914 \tabularnewline
\hline
\end{tabular}
\end{table}

The results show that, in both PTP and RTP cases, interest similarity among females is on average larger than that among males. One possible explanation is that females' interest is more concentrated leading to a higher probability to share interests between females. To check this conjecture, we randomly selected the same number of males and females, and found that the tag set for female group is smaller. This means that females have a narrower range of video interests compared to males. 

\subsubsection{Age.}
We filtered out users of age greater than 40 or less than 10 since there are few users in the two ranges.
One intuition is that users with close ages may be more alike. Thus, we first explored the relation between interest similarity and age differences between any two users. However, almost no distinction in the interest similarity was observed over different age differences.

\begin{figure}[htb]
  \centering
  \subfigure[]{
     \includegraphics[width=1.6in]{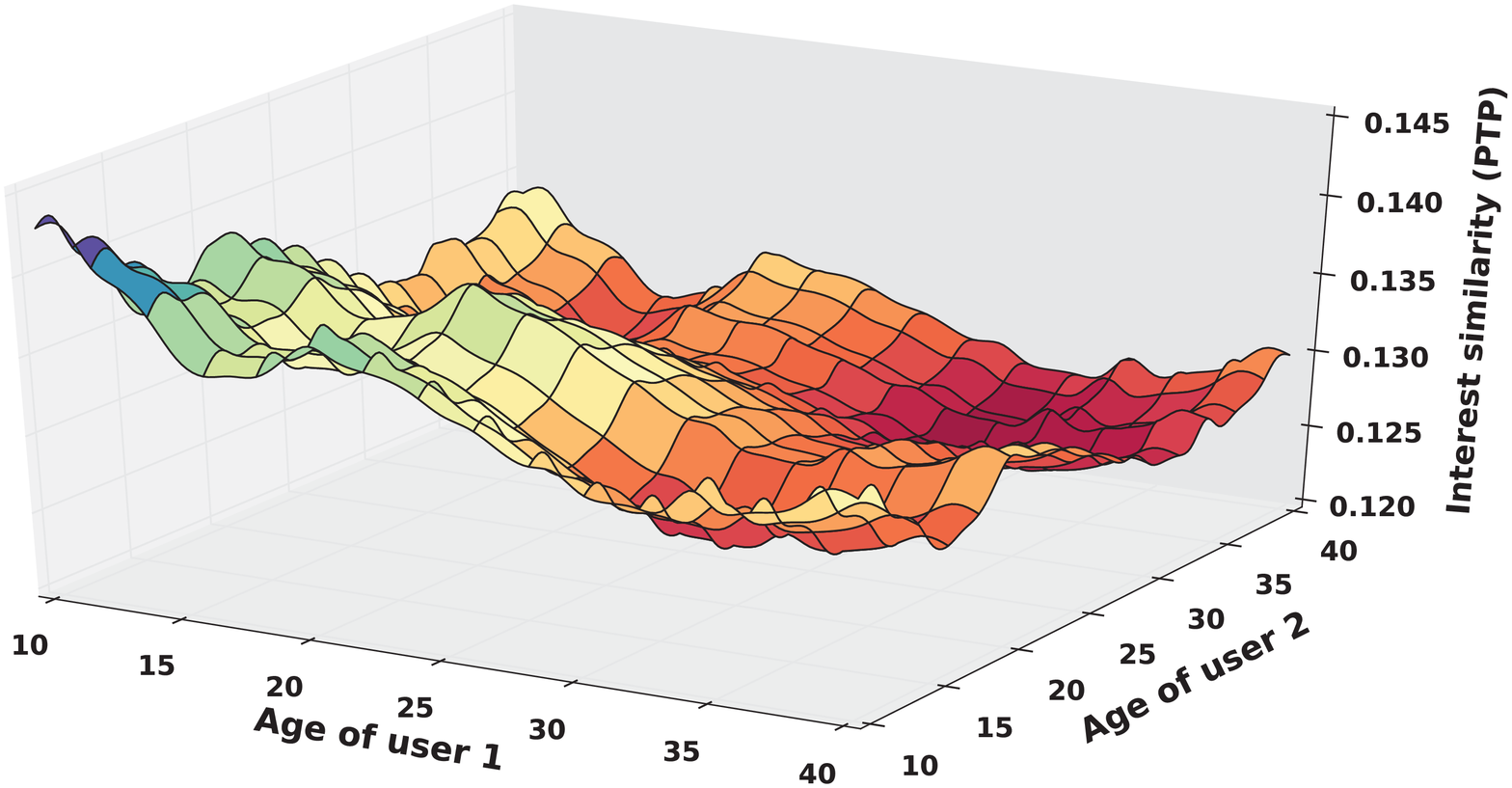}}
  \subfigure[]{
		 \includegraphics[width=1.6in]{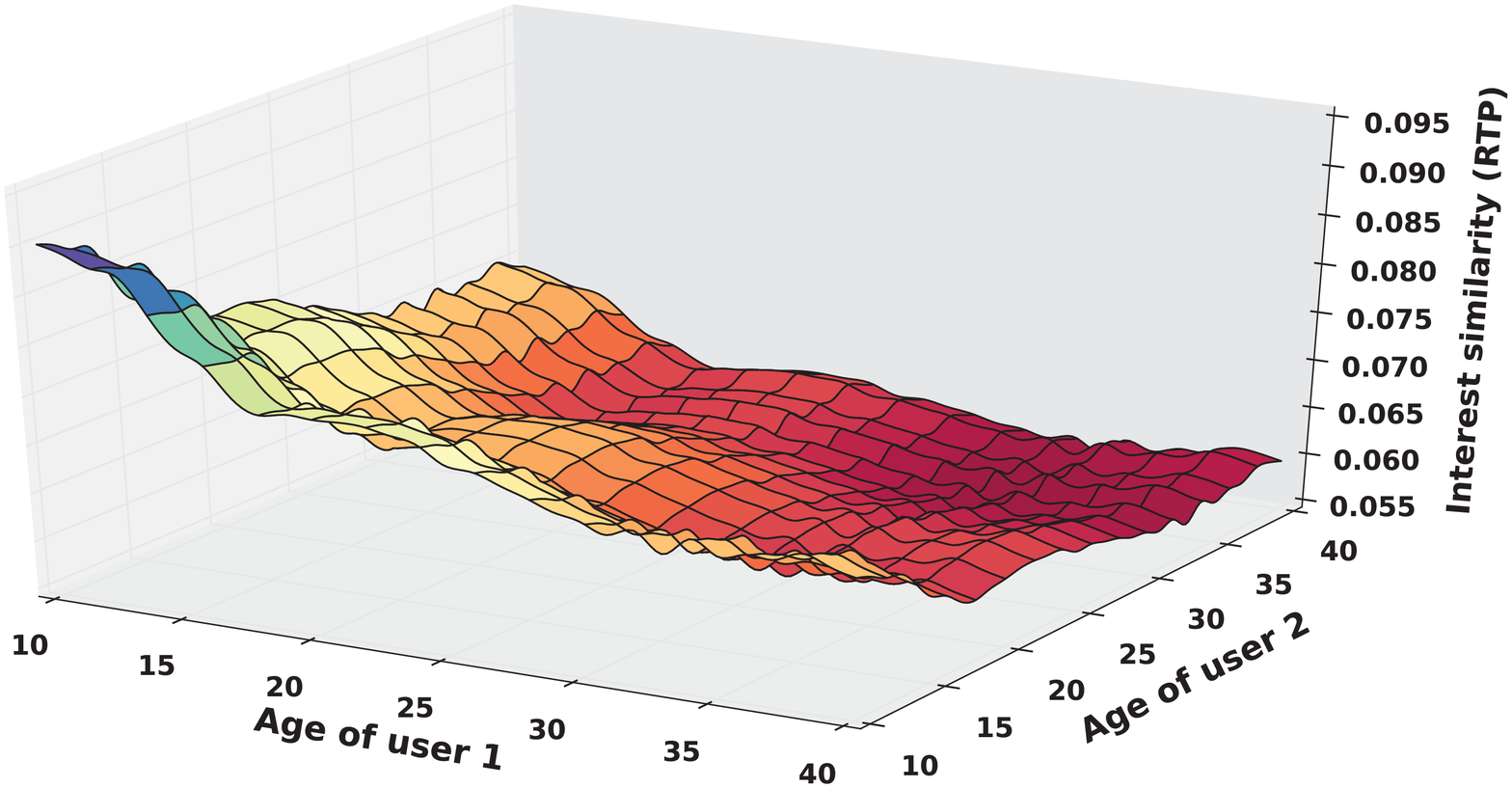}}
	\caption{Interest similarity between users in each age pair.}
  \label{Fig:sim_vs_age_3d} 
\end{figure}

It is possible that users in various generations differ in their interest similarity. For example, two users of age 15 and 16 has the same age difference as that for users of age 49 and 50, but the interest similarity may be not the same. Therefore, we further explored the interest correlation of age-age pairs instead of age differences. To illustrate the measured interest similarity, we draw three-dimensional plots as shown in Fig.~\ref{Fig:sim_vs_age_3d}.

Each point in Fig.~\ref{Fig:sim_vs_age_3d} is the average interest similarity between users with the pair of ages. From the results, we can observe that age pair as a feature contains more information than the age difference, that is, age pair is more discriminative in similarity inference\footnote{For the sake of statistical significance, we drew a large number of user pairs for each point. Similar treatment was adopted in the other parts of the measurement.}.

\subsubsection{Location.}
Intuitively, people from the same place may be interested in similar videos, such as local news. However, from the empirical results, we found that the average similarity between users from the same city is not higher than random cases. This implies weak geographical limitation for user video interests, while the QQ friendship shows stronger locality, because in our dataset, the probability of two friends from the same city is 18 times as high as that of two random users. For the purpose of interest similarity prediction, location pairs, as we measured, will be more useful than the binary feature, i.e., whether two users are from the same city or not. 

\subsection{Correlation of Social Relationship to Their Interest}
Studies of social influence and homophily~\cite{lewis2012social} indicate that members with social relationship often exhibit correlated behaviors and similar interests. However, the correlation between social relations and interests is not intuitive because social relations are quite heterogeneous. In this paper, we investigated different aspects of social information in terms of 1) different strength of friendship; 2) direct friendship and indirect friendship; 3) friendship and group membership based on QQ and QQ group information. 
\subsubsection{Friendship.}
We first examined direct friendship in general.
Results in Table~\ref{tab:friendship} show that two friends appear to be slightly more similar than a pair of strangers, but the difference is not very obvious.
\begin{table}[htb]
\centering
\caption{Average interest similarity vs. location of two users}\label{tab:friendship}
\begin{tabular}{|c|c|c|}
\hline
 & PTP  & RTP \tabularnewline
\hline
Friends & 0.14  &   0.072 \tabularnewline
\hline
Random  & 0.1308 &  0.0658 \tabularnewline
\hline
\end{tabular}
\end{table}

\subsubsection{Interaction Intensity and Frequency.}
\begin{figure}[htb]
  \centering
  \subfigure[]{
     \includegraphics[width=1.6in]{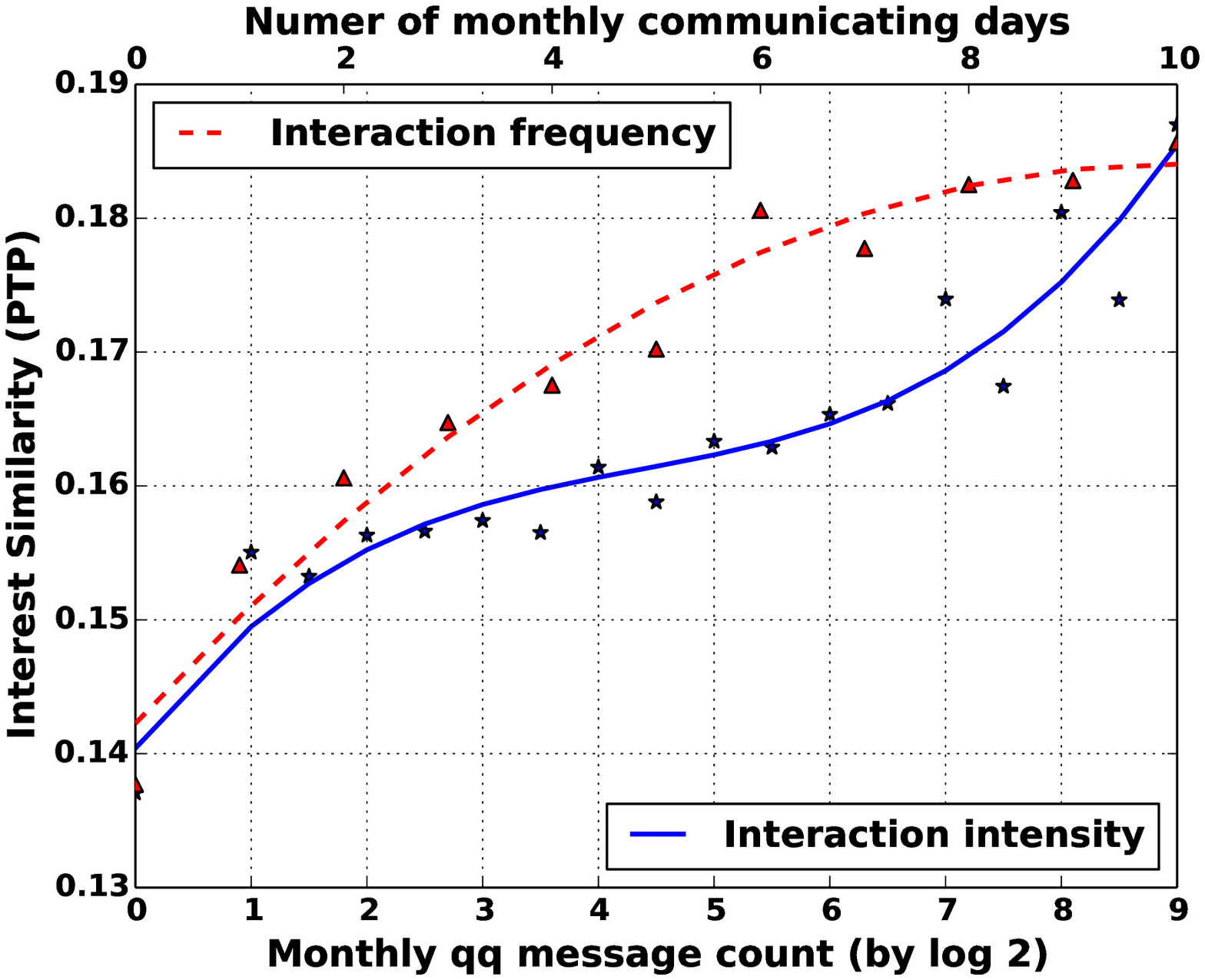}}
  \subfigure[]{
		 \includegraphics[width=1.6in]{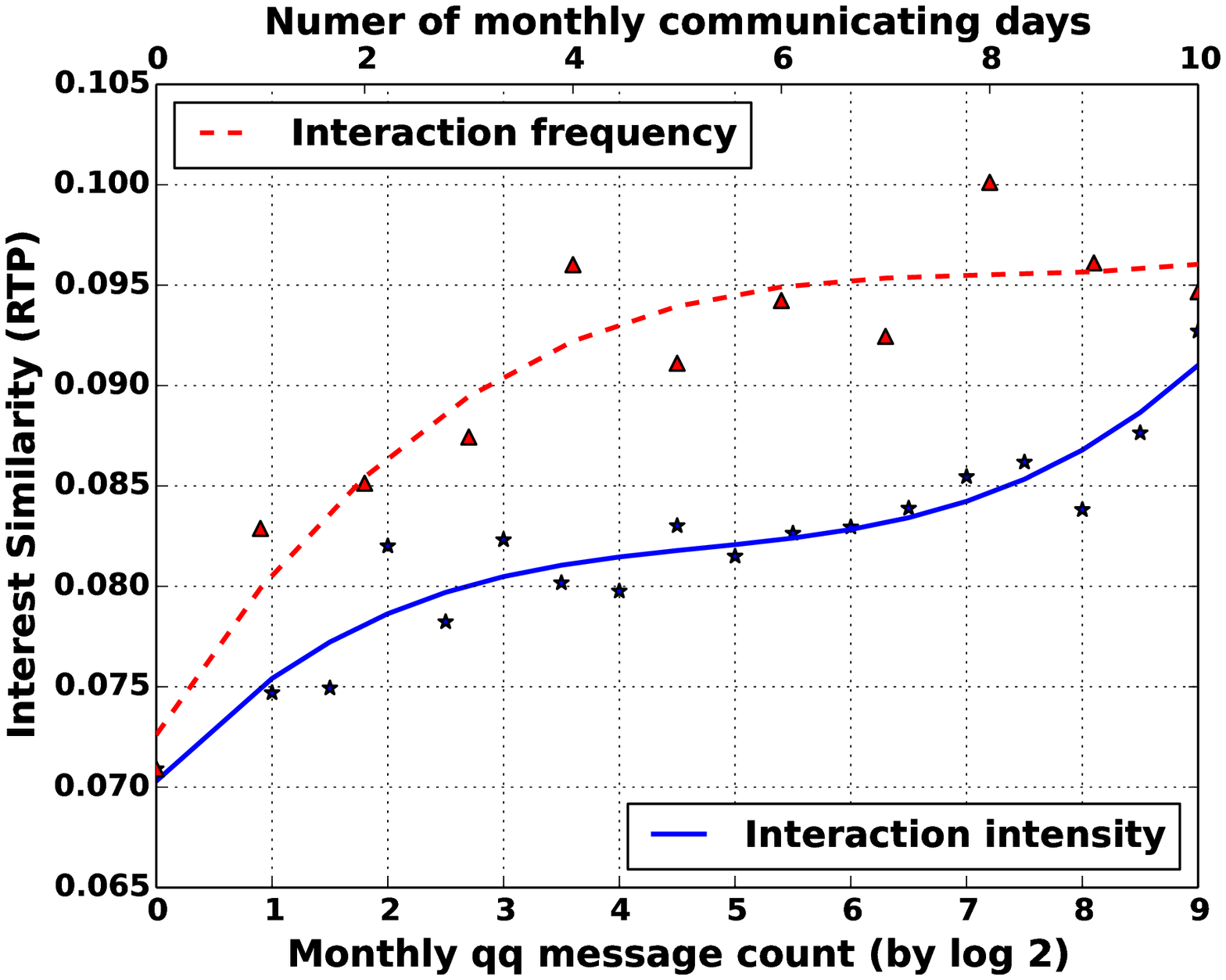}}
	\caption{Interest similarity vs. monthly qq message count and number of monthly communicating days.}
  \label{Fig:sim_vs_msg_cnt_and_days} 
\end{figure}

For friendship, some are close friends, and some are merely acquaintances. 
One way to measure the strength of friendship is based on interaction intensity and frequency~\cite{petroczi2007measuring}. 
In our system, interaction intensity is measured by monthly total message count between two users, denoted by $m_{-30:-1}^C(u,v)=\sum_{d=-30}^{-1}{m_d(u,v)}$, while the interaction frequency means the number of days two users communicated with each other in the past one month, represented by $m_{-30:-1}^D(u,v)=\bigl|{d|d \in [-30,-1], m_d(u,v)>0}\bigr|$. 
We calculated the average similarity of friends with different interaction intensity and interaction frequency, as shown in Fig.~\ref{Fig:sim_vs_msg_cnt_and_days}. Although the fitted curves\footnote{We utilized fitting curves to show the trends clearly, and we did not show functions of the curves here because we do not need the exact functions in the interest similarity prediction.} in the RTP case is not the same as that for PTP, they both show that users with higher intensity of interaction share more interests, particularly when the interaction intensity is very large. Also, interest similarity increases more sharply when number of communicating days is small. 
Interestingly, interaction frequency is more correlated with interest similarity than interaction intensity, implying that some casual or transactional interaction could also have a large intensity. 

\subsubsection{Common Friendship.}
Besides direct friendship, indirect friends may also have influence. In our study, like the phenomenon of triadic closure in social network~\cite{kossinets2006empirical}, we only considered the relationship with two hops\footnote {We did not talk about the relationship with more than two hops because the observed interest similar was very weak.} in the OSN.

\begin{figure}[htb]
  \centering
  \subfigure[]{
     \includegraphics[width=1.6in]{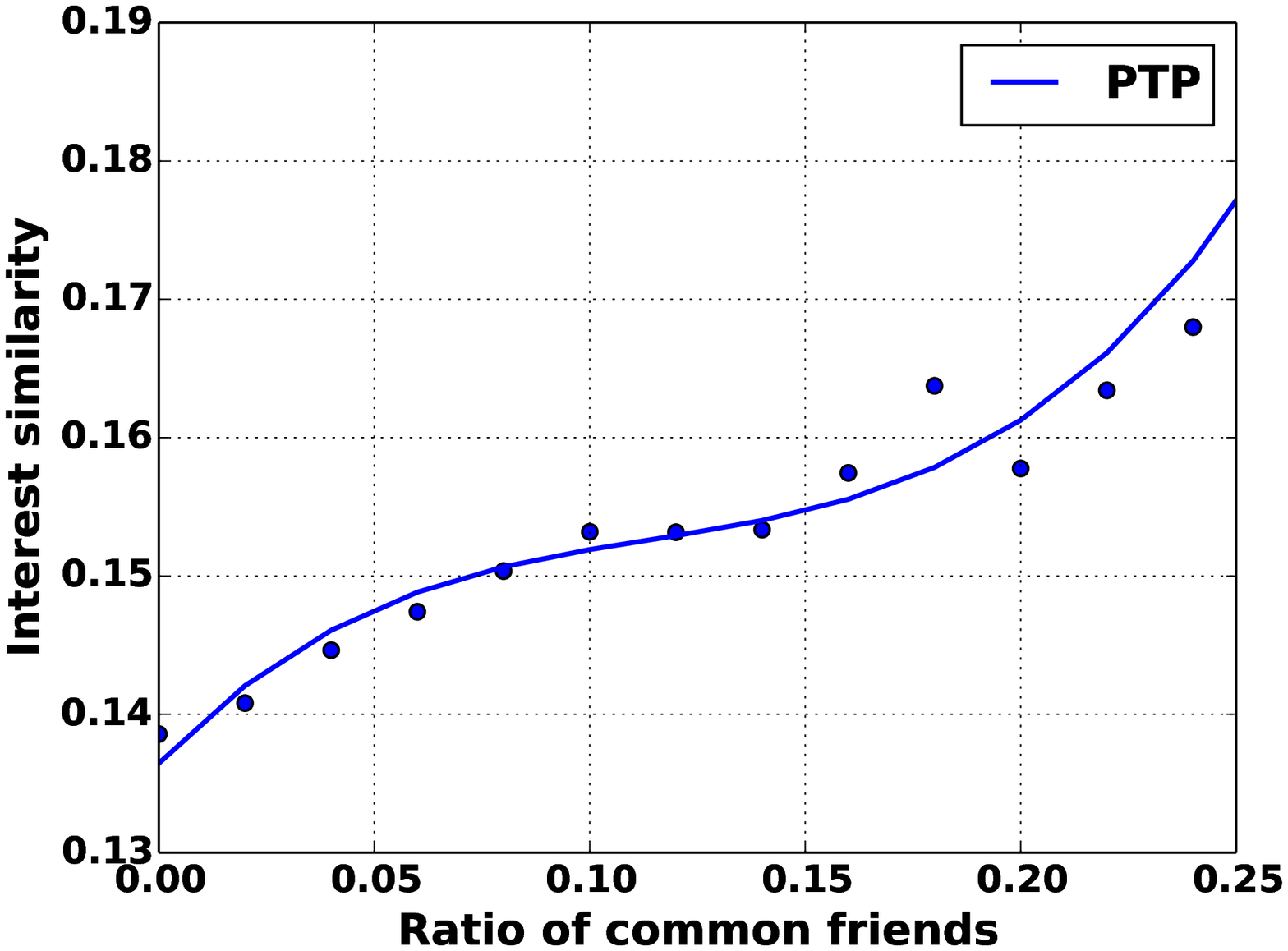}}
  \subfigure[]{
		 \includegraphics[width=1.6in]{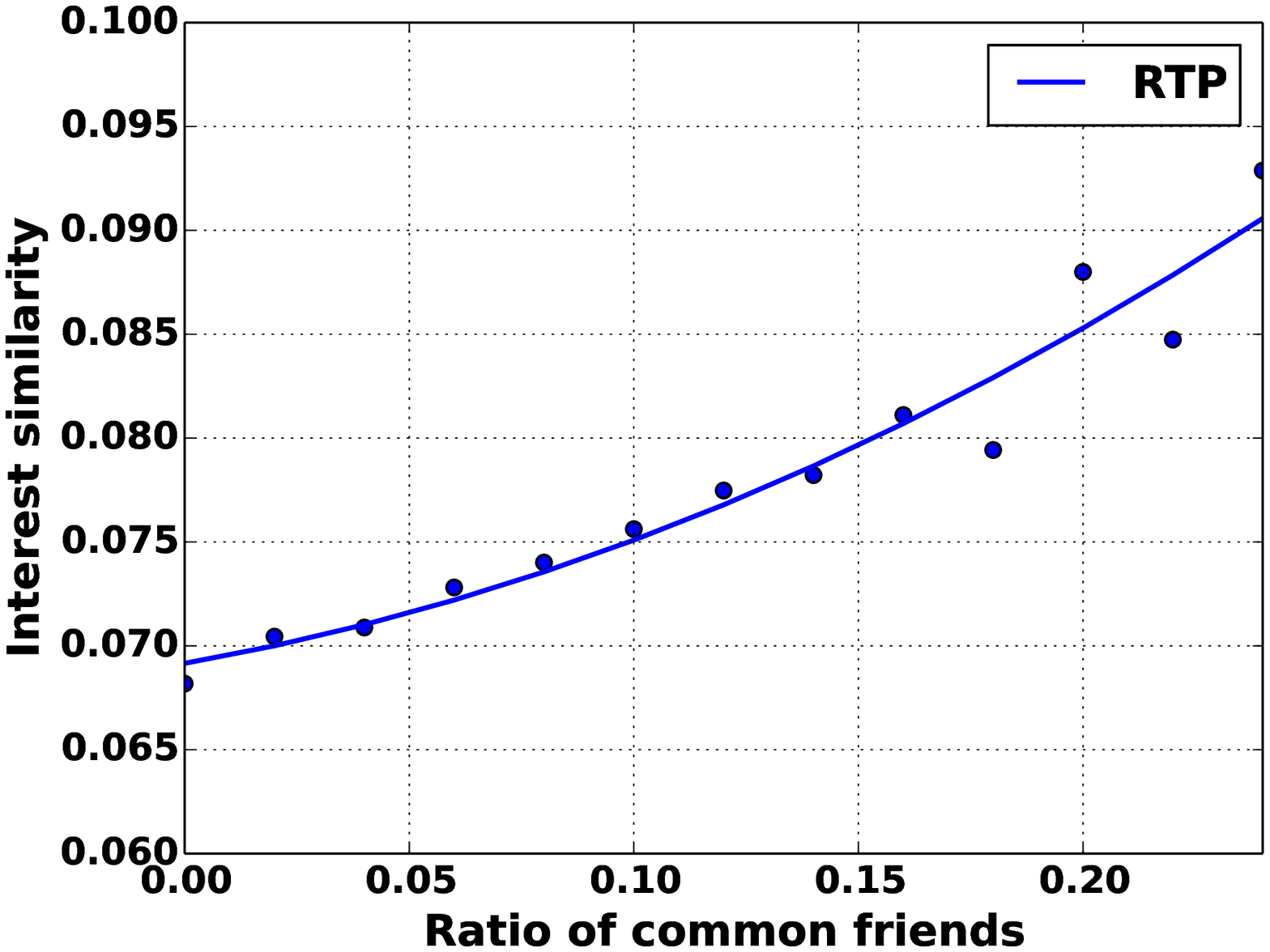}}
	\caption{Interest similarity vs. ratio of common friends.}
  \label{Fig:sim_vs_common_friend_ratio} 
\end{figure}

We firstly investigate the correlation between the number of common friends between user $u$ and $v$:$|\mathcal{F}_u \cap \mathcal{F}_u|$ and interest similarity.
However, we did not observe a significant relation between interest similarity and the number of common friends. In fact, it's more likely to have a common friend for two users with many friends. Thus, overlap of two users' friend sets should be normalized by the size of each set, i.e., $\frac{\left|\mathcal{F}_u\cap \mathcal{F}_v\right|}{\sqrt{\left|\mathcal{F}_u\right|}*\sqrt{\left|\mathcal{F}_v\right|}}$. For most user pairs, the ratio of common friends is less than 0.25. The correlation between interest similarity and the ratio of common friends is shown in Fig.~\ref{Fig:sim_vs_common_friend_ratio}.

The results in Fig.~\ref{Fig:sim_vs_common_friend_ratio} indicate that users with larger ratio of common friends have more similar interests.

\subsubsection{Common Group Membership.}
By intuition, common group affiliation indicates similar interest. Hence, we examined the correlation between the number of common groups between user $u$ and $v$, i.e., $|\mathcal{G}_u \cap \mathcal{G}_v|$, and their interest similarity. 

However, no discernable correlation can be observed between interest similarity and the number of common groups. And even after we normalized the number of common groups by number of groups joined by each user, we could not observe a clear correlation yet.

\begin{figure}[htb]
  \centering
  \subfigure[]{
     \includegraphics[width=1.6in]{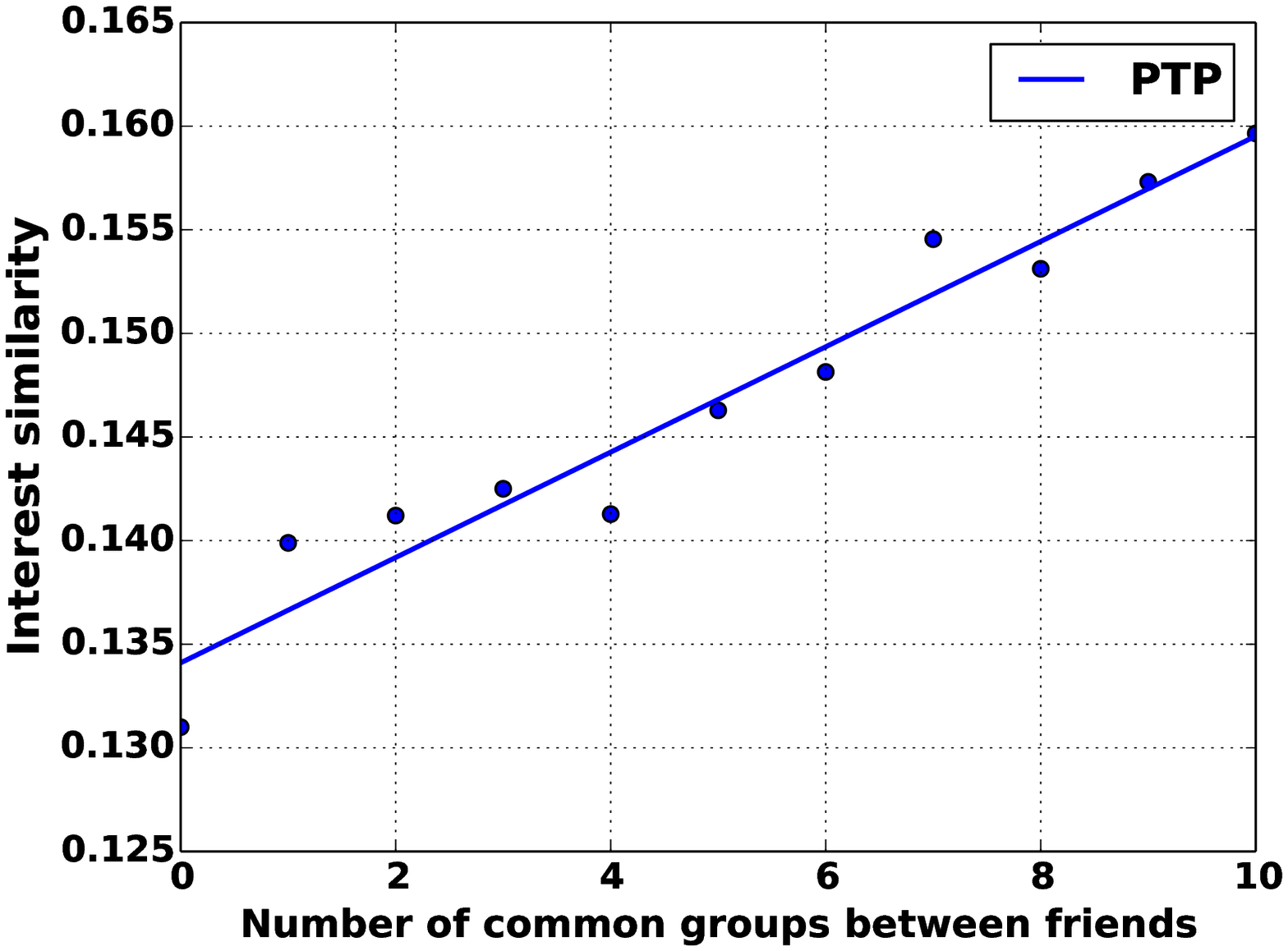}}
  \subfigure[]{
		 \includegraphics[width=1.6in]{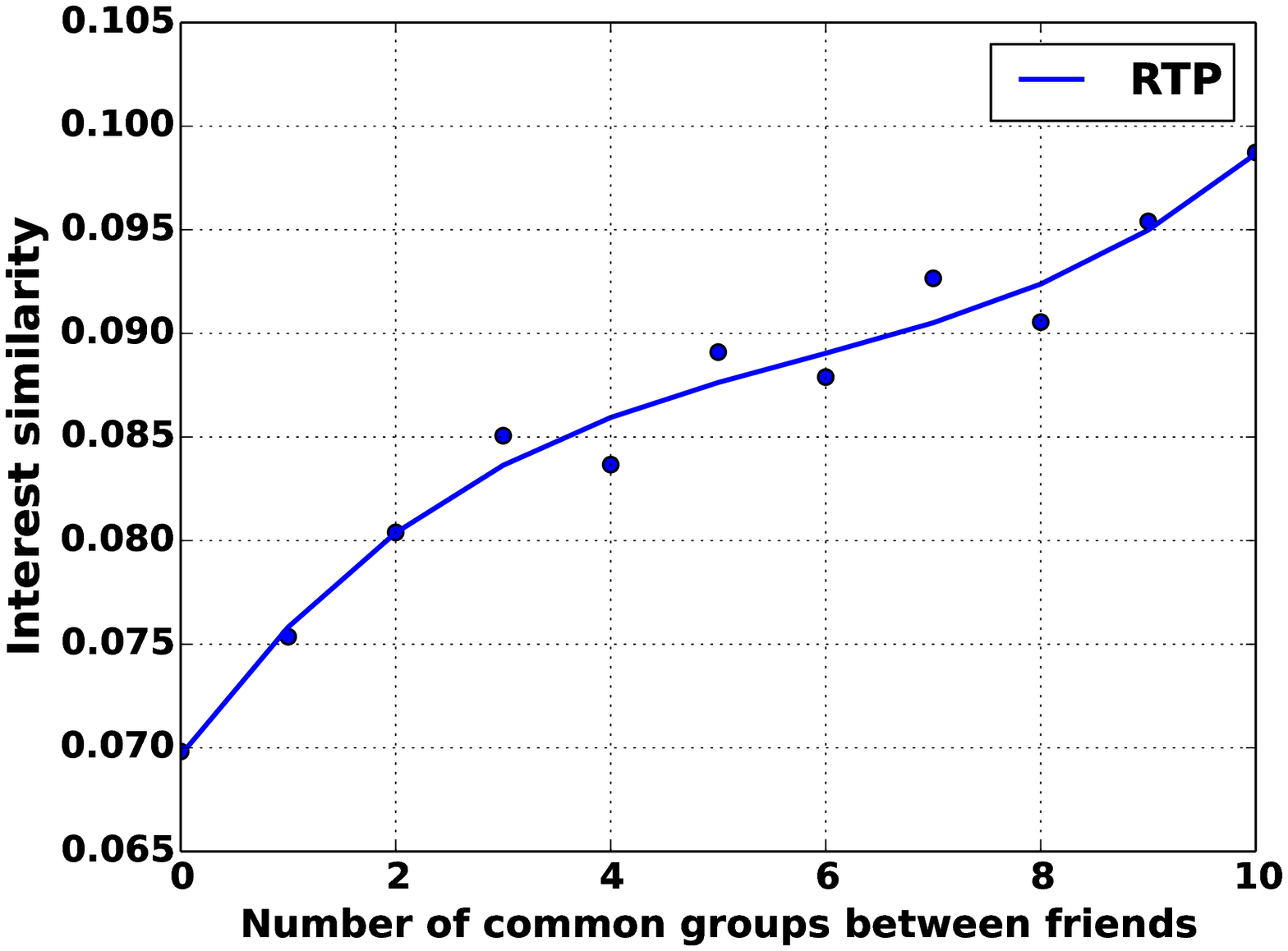}}
	\caption{Interest similarity vs. number of common groups between friends.}
  \label{Fig:sim_vs_common_group_and_friendship} 
\end{figure}

This implies that group membership is weakly related to the interest similarity. But it is more effective if the membership is combined with friendship, as shown in Fig.~\ref{Fig:sim_vs_common_group_and_friendship}. This may result from the previous empirical observations that social relationship and interests are generically confounded~\cite{shalizi2011homophily}. Users may become friends due to common interests, while social influence between friends may also enhance the interest conformity.

\subsection{Interest Evolution Over Time}
\subsubsection{Interest Evolution.}
\begin{figure}[htb]
  \centering
  \subfigure[]{
     \includegraphics[width=1.6in]{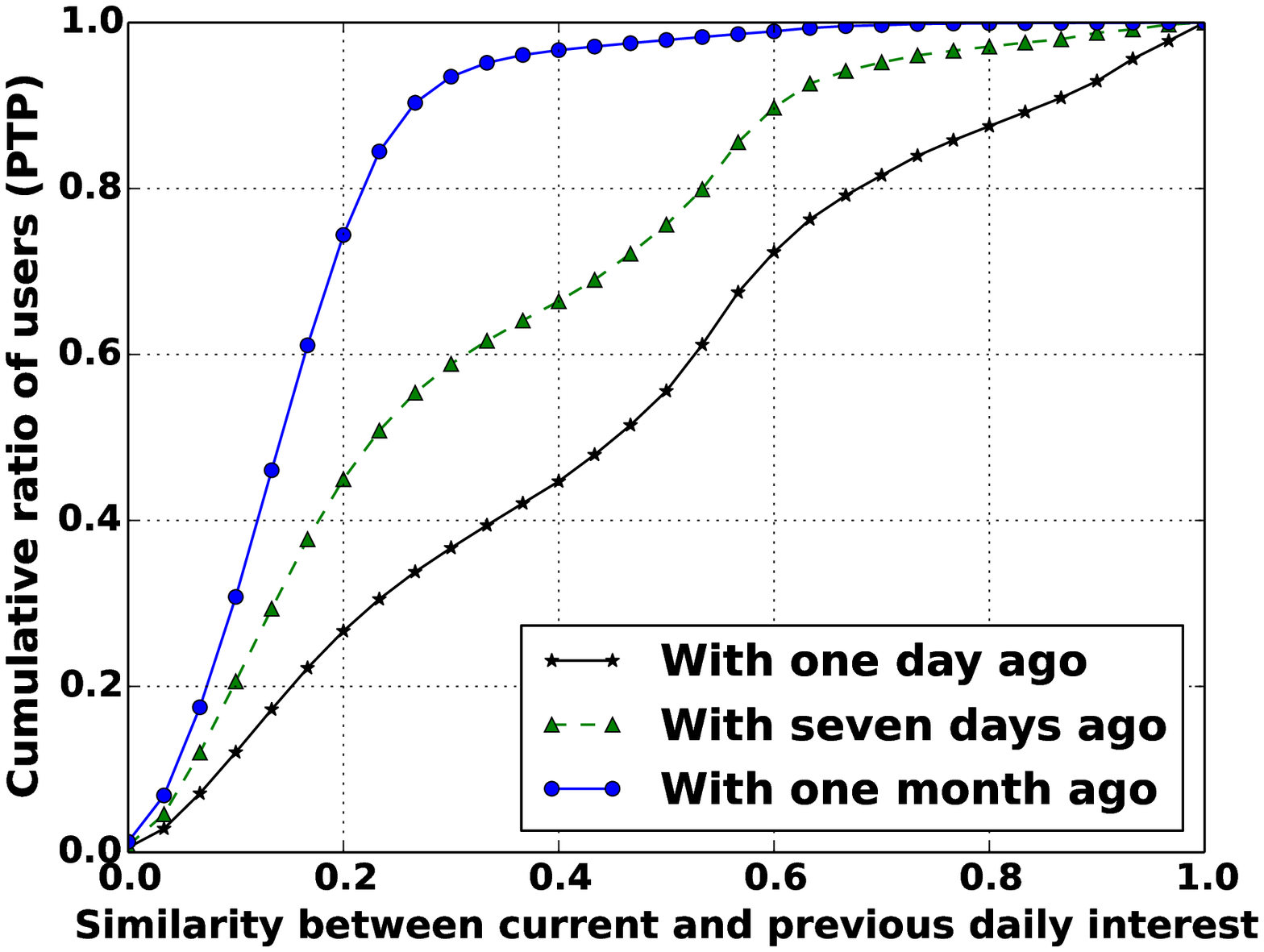}}
  \subfigure[]{
		 \includegraphics[width=1.6in]{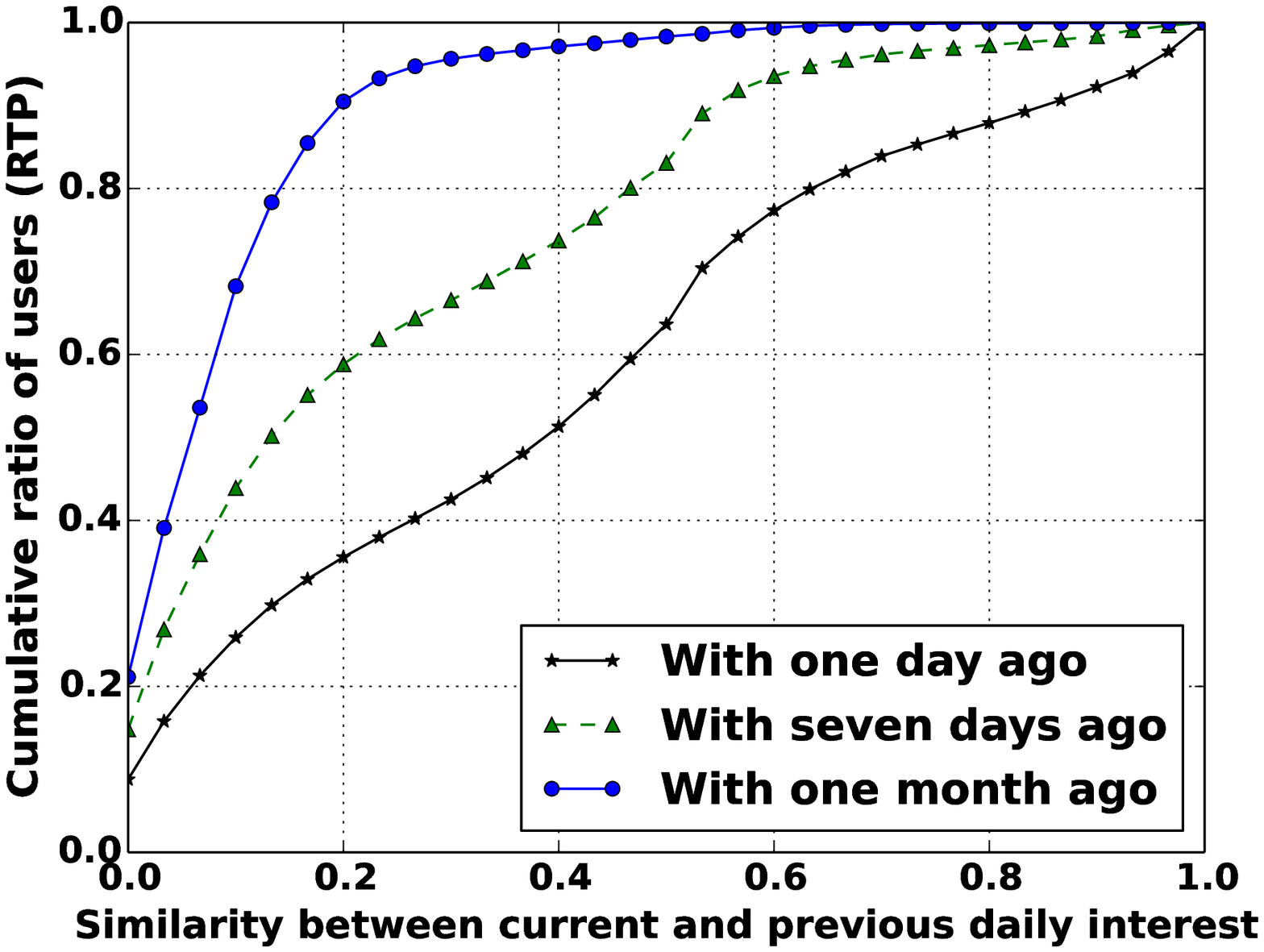}}
	\caption{Distribution of self-similarity of interest between different days.}
  \label{Fig:interest_evolve} 
\end{figure}

We compared the current daily tag-based profiles with different previous daily profiles for the same user, and the results in Fig.~\ref{Fig:interest_evolve} show that users' interests deviated more from the current ones as the time gap expanded in both PTP and RTP. In general, there are three potential causes of the above observations: 1) interests evolved; 2) topical trends shifted; 3) interests were partially expressed by behaviors.
The influence of the second factor could be revealed using the RTP method. With massive data on each day, the third factor is uniform across the tested days, and thus won't affect the distributions. Therefore, on the basis of the observations under both PTP and RTP methods, we argued that there exists user interest evolution. This is also the reason to study and predict current interest similarity. Note that the phenomenon of interest evolution cannot be illustrated under the video-based interest profiling.

\subsubsection{Coverage and Correlation.}
When there is little or no recent behavioral data, an intuitive idea is to explore the past long-term data. For users who are currently inactive, they might have behaviors some time ago, such as in the past one week, past one month, etc. Basically, the longer the duration, the more users would have some behavior records, i.e., higher coverage of users. According to the system measurement, among the users who are active on the current day, 27.5\% of them were also active in the past one day; around 67\% of them were active in the past one week; and more than 85\% of them had some behaviors in the past one month. However, as shown before, users' interests evolved with time. Interests induced from behaviors long time ago may deviate a lot from the current interests i.e., lower correlation with current interest. Therefore, we studied the correlation between the current interest similarity between users and their past similarity with different durations. Similar to notations of social interaction, we denote interest similarity in the past one day, past one week, and past one month as $S_{-1}^P(u,v)$, $S_{-1}^R(u,v)$; $S_{-7:-1}^P(u,v)$, $S_{-7:-1}^R(u,v)$; and $S_{-30:-1}^P(u,v)$, $S_{-30:-1}^R(u,v)$, respectively.
The Pearson correlation coefficient between interest similarity on the current day and that in the past duration is in Table~\ref{tab:corr_vs_coverage}.
\begin{table}[htb]
\centering
\caption{correlation between interest similarity on the current day and that in the past duration}\label{tab:corr_vs_coverage}
\begin{tabular}{|c|c|c|}
\hline
 & PTP  & RTP \tabularnewline
\hline
With past one day & 0.36  &   0.34 \tabularnewline
\hline
With past one week  & 0.28 &  0.27 \tabularnewline
\hline
With past one month & 0.22  &   0.19 \tabularnewline
\hline
\end{tabular}
\end{table}

\subsection{Individuality of Interest}
We propose the individuality of interest which indicates how much a user is similar to other users on average. On the basis of formulas of cosine similarity measure in Eq.~\ref{eq:PTP_sim} and Eq.~\ref{eq:RTP_sim}, individuality is affected by three factors: 1) probability of sharing a common tag; 2). weights of the common tags; 3). norm of the user profile's weight vector. While the first two factors influence the numerator of the formulas, the third factor affects the denominator. Thus, we defined the individuality of user $u$ as
\begin{eqnarray}
\begin{aligned}
H^P\left(u\right)=\frac{\sum_{i\in \mathcal{T}_u}\mathbb{T}_u^P[i]*|\mathcal{U}_i|}{\sqrt{\sum_{i\in \mathcal{T}_u}(\mathbb{T}_u^P[i])^2}*|\mathcal{U}|}
\end{aligned}
\end{eqnarray}
\begin{eqnarray}
\begin{aligned}
H^R\left(u\right)=\frac{\sum_{i\in \mathcal{T}_u}\mathbb{T}_u^R[i]*|\mathcal{U}_i|}{\sqrt{\sum_{i\in \mathcal{T}_u}(\mathbb{T}_u^R[i])^2}*|\mathcal{U}|}
\end{aligned}
\end{eqnarray}

For the first factor, the probability of both two users have a certain tag is determined by the global popularity of this tag, namely, the number of users owning this tag ($|\mathcal{U}_i|$). When two users share a tag, the weight of this tag, i.e., $\mathbb{T}_u^P[i]$  or $\mathbb{T}_u^R[i]$ in each user's profile, decides the contribution to the numerator. $|\mathcal{U}|$ is used to normalize the feature.

\begin{figure}[htb]
  \centering
  \subfigure[]{
     \includegraphics[width=1.6in]{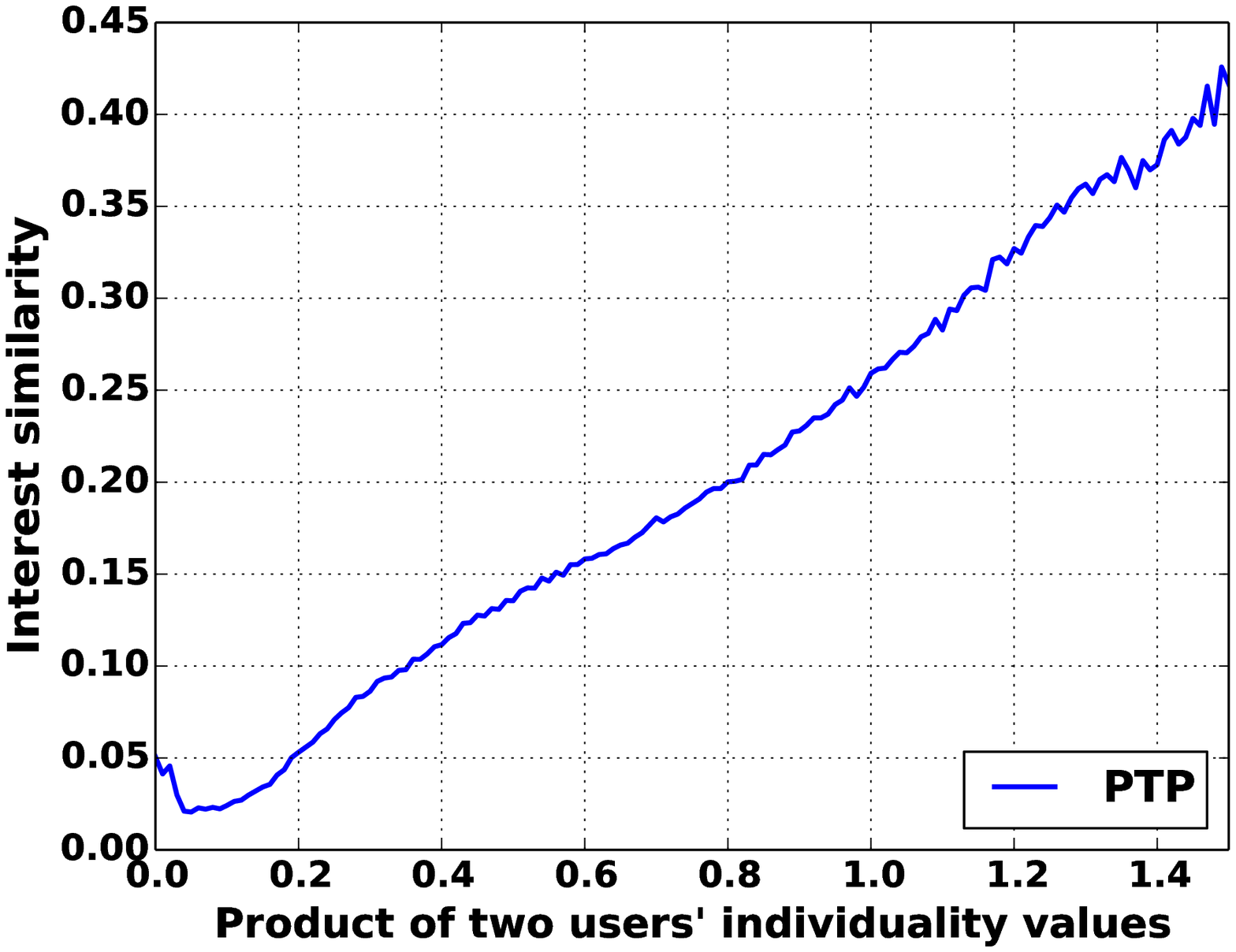}}
  \subfigure[]{
		 \includegraphics[width=1.6in]{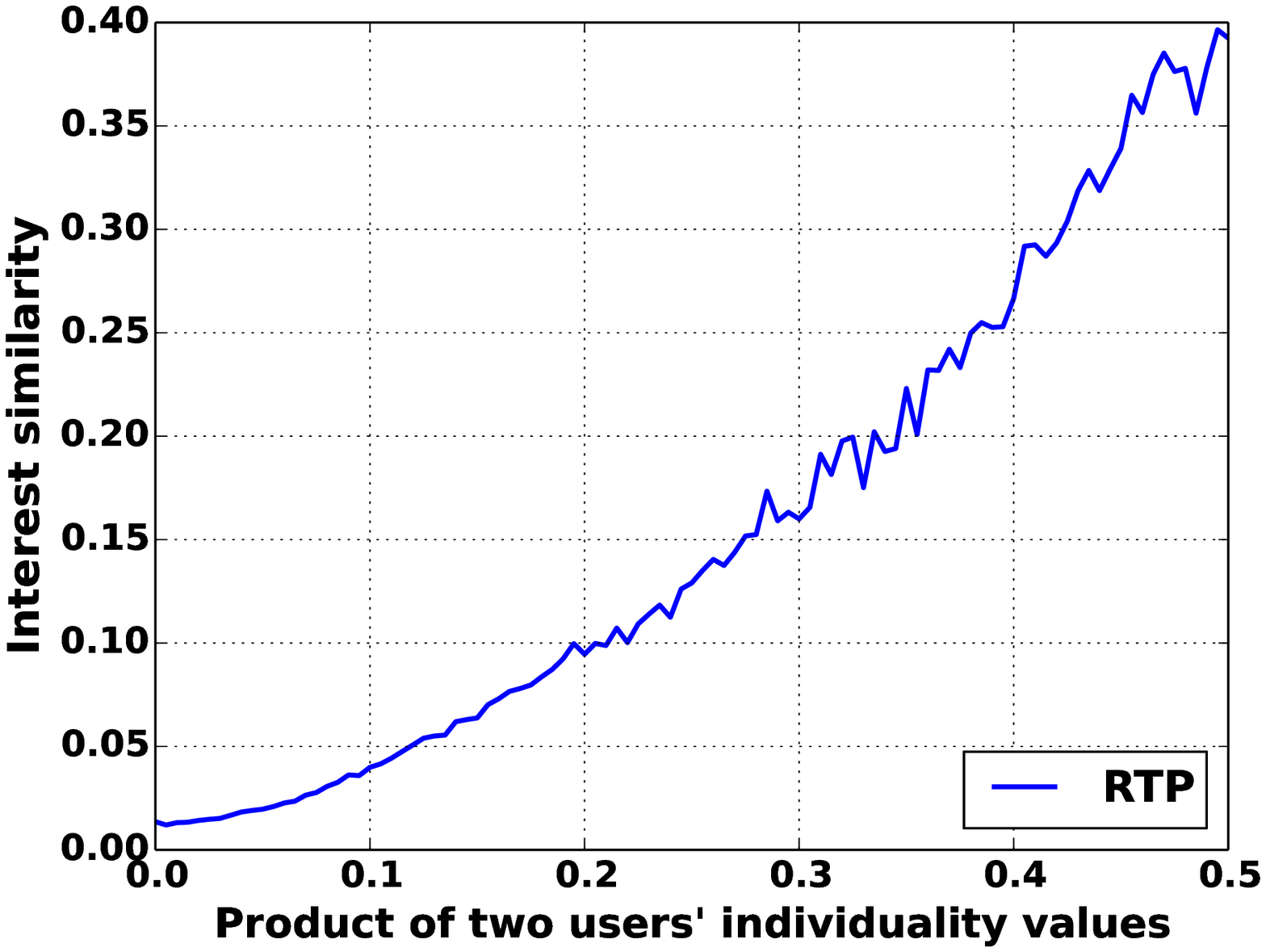}}
	\caption{Interest similarity vs. product of two users' individuality.}
  \label{Fig:sim_vs_individuality_product} 
\end{figure}

We randomly selected one million pairs of users. Then we compared the interest similarity between two users and the product of their individuality values, as shown in Fig.~\ref{Fig:sim_vs_individuality_product}. We observed that two users with large individuality values are more similar in general. The individuality of interest can be used as an effective feature for similarity inference if we know the interests of one user in a user pair.

\section{Inferring Interest Similarity Between Users}
\label{Sec:simPrediction}
The goal of this section is to infer user interest similarity based on the observations in Sec.~\ref{Sec:measurement}. From the empirical studies, we obtained the following insights: 1) we need to pre-process features properly to make them more discriminative, such as normalizing the number of common friends; 2) not all the relations are quasi-linear, such as relations between interaction intensity and interest similarity; 
3) some features need to be combined to make difference in inferring interest similarity, such as membership combined with friendship. 

In this section, we attempted to investigate two issues: 1) among different machine learning models, which one is best suited for our prediction tasks; 2) 
the effectiveness of different categories of features and their combinations.
To illustrate them, we conduct two series of experiments: 1) with all the selected features, we fit them with various algorithms to learn different models; 2) for each feature combination, we construct a predictive model using pruned decision tree. For each series of experiments, we examine the performance of  interest similarity prediction in both PTP and RTP cases for the classification (\textbf{two users are similar or not}) and regression tasks (\textbf{the similarity value between two users}).

In total, we define 10 features belonging to three categories, namely demography, social relations and interest:
\begin{multicols}{2}
  \begin{itemize}
\item Gender pair: $g_u$ and $g_v$
\item Age pair: $a_u$ and $a_v$
\item Location pair: $l_u$ and $l_v$
\item Friendship: $F(u,v)$
\end{itemize}
\end{multicols}
\begin{itemize}
\item Ratio of common friends: $\frac{\left|\mathcal{F}_u\cap \mathcal{F}_v\right|}{\sqrt{\left|\mathcal{F}_u\right|}*\sqrt{\left|\mathcal{F}_v\right|}}$
\item Number of common groups: $|\mathcal{G}_u \cap \mathcal{G}_v|$
\item Monthly message count: $m_{-30:-1}^C(u,v)$
\item Monthly messaging days: $m_{-30:-1}^D(u,v)$
\item Past long-term similarity: $S_{-30:-1}^P(u,v)$ or $S_{-30:-1}^R(u,v)$
\item Individuality: $H^P\left(u\right)$ or $H^R\left(u\right)$
\end{itemize}

From users who were active on the target day (August 30th, 2015), we randomly selected one million user pairs for each experiment. We used seventy percent of the samples for training and the rest as the testing data. Moreover, ten-fold cross validation was used in model training.
For the classification task, we binarized the similarity values with a certain threshold (the mean value of interest similarity) so that we could predict whether two users are similar or not. To evaluate the performance, we utilized the evaluation metrics of area under the ROC curve (\textbf{AUC}) which is insensitive to label imbalance. And for regression, to avoid the specific values of regression error, we used reduced ratio of mean absolute error (\textbf{reduced ratio of MAE}) as the performance metrics which is calculated by comparing the error with that achieved by a constant estimator using the mean value of similarity in the training set.

\subsection{Experiments of Various Machine Learning Models }
Particularly, we are interested in predicting the interest similarity between inactive users and active users since it's useful in applications, such as item recommendation~\cite{su2009survey} and targeted online advertising~\cite{yu2014facebook}. 
Since we have the interest information of the active user, we could utilize her interest individuality as a predictive feature\footnote{If both users are currently inactive, then we only need to omit the individuality feature in the prediction.}.

The next job is to select the best model to deal with these features.
Considering the heterogeneity of these features, it is intuitive to use tree models which are non-linear models and can cope with categorical and continuous features directly; 
while linear models are better at approaching the data on the whole and fitting linear or quasi-linear relations. To capitalize on the merits of both models, we propose a hybrid tree-encoded linear model. In this model, we firstly used gradient boosting decision tree (GBDT)~\cite{ye2009stochastic} to encode the features by converting the original features into some binary features. Each encoded feature corresponds to a region jointly described by multiple ranges of original feature values. As a simple example, consider a boosted tree with two sub-trees, where the first sub-tree has three leaves and the second has two leaves. If an input sample ends up in leaf two in the first sub-tree and leaf one in second sub-tree, then the input will be encoded as the binary vector [0,1,0,1,0], where the first three entries correspond to the leaves of the first sub-tree and last two to those of the second sub-tree. The details of using GBDT to encode heterogeneous features into binary features can be found in~\cite{he2014practical}. Furthermore, to complement the possibly missed linear relations between the output and the original features, we applied both the encoded features and original features into a regularized linear model, i.e., logistic regression with L-1 regularization for classification and Lasso regression for regression, to reduce redundancy among those features. 

For comparison, we used two linear models and two tree models to fit the data. The linear models consist of simple linear models and regularized linear models, and the tree models comprise decision tree models and ensemble tree models. For linear models, regularization can be embedded to alleviate the influence of feature dependence; while for tree models, ensemble~\cite{alpaydin2014introduction} is a state-of-the-art approach widely used to improve the performance. In our experiments, we utilized linear regression and logistic regression as well as the l1-regularized ones of them for the classification and the regression, respectively. Moreover, pruned decision tree~\cite{liu2010robust} and random forest\footnote{The individual performance of random forest and gradient boosting decision tree (GBDT) is almost the same, thus we only show one of them.}~\cite{saffari2009line} were used for both classification and regression tasks. 
The performance results of different models in similarity prediction are shown in Table~\ref{tab:classification_diff_alg} and Table~\ref{tab:regression_diff_alg}.

\begin{table}[htb]
\centering
\caption{The performance of different models in the classification task}\label{tab:classification_diff_alg}
\begin{tabular}{|c|c|c|}
\hline
Metrics: AUC & PTP  & RTP \tabularnewline
\hline
Logistic regression & 0.783  &   0.776 \tabularnewline
\hline
L1-Regularized Logistic regression  & 0.805 &  0.796 \tabularnewline
\hline
Pruned decision tree  & 0.820  &   0.810 \tabularnewline
\hline
Random forest & 0.828 &   0.818 \tabularnewline
\hline
Hybrid tree-encoded linear model & \textbf{0.834}  &  \textbf{0.826} \tabularnewline
\hline
\end{tabular}
\end{table}

\begin{table}[htb]
\centering
\caption{The performance of different models in the regression task}\label{tab:regression_diff_alg}
\begin{tabular}{|c|c|c|}
\hline
Metrics: reduced ratio of MAE & PTP  & RTP \tabularnewline
\hline
Linear regression & 18.33\%  &   16.14\% \tabularnewline
\hline
Lasso regression  & 19.23\% &  17.36\% \tabularnewline
\hline
Pruned decision tree  & 23.88\% &  21.28\% \tabularnewline
\hline
Random forest & 24.5\% & 21.69\% \tabularnewline
\hline
Hybrid tree-encoded linear model & \textbf{25.5\%}  &  \textbf{23.0\%} \tabularnewline
\hline
\end{tabular}
\end{table}

Under both PTP and RTP, the hybrid tree-encoded linear model achieved the best performance in the classification and regression tasks, meaning that it best fit our problem. In fact, the tree-encoded features could achieve feature combinations automatically so as to capture the non-linear and multi-feature (i.e., membership combined with friendship) relations. Different from traditional ensemble tree models, namely random forest and GBDT which assigns a weight for each sub-tree, our hybrid tree-encoded linear model will learn a weight for each leaf node of the sub-trees and for each original feature.

We also applied the hybrid tree-encoded linear model to predict video-based similarity defined in Eq.~\ref{eq:vid_sim} with our full feature set, and compared the results with that using the tag-based scheme as shown in Table~\ref{tab:PTP_rtp_vbp}. The predicted video-based similarity will be used in Sec.~\ref{Sec:recommendation}.

\begin{table}[htb]
\centering
\caption{The performance of our model using different user profiling methods and all the features}\label{tab:PTP_rtp_vbp}
\begin{tabular}{|c|c|c|}
\hline
  & AUC  & reduced ratio of MAE \tabularnewline
\hline
PTP & 0.834  &  25.5\% \tabularnewline
\hline
RTP  & 0.826 &  23.0\%  \tabularnewline
\hline
Video-based profiling & 0.792 &  11.7\%  \tabularnewline
\hline
\end{tabular}
\end{table}


\subsection{Experiments of Different Feature Combinations}
To test the predictive ability of different features, we can use seven combinations, namely, social, demographic, interest, social + demographic, social + interest, demographic + interest, and social + demographic + interest. 
For each combination, we utilize pruned decision tree to learn the model, since it can capture non-linear patterns and achieve good performance with careful pruning. The results are shown in Table~\ref{tab:classification_diff_feature} and Table~\ref{tab:regression_diff_feature}.

\begin{table}[htb]
\centering
\caption{The performance of different feature combinations in the classification task}\label{tab:classification_diff_feature}
\begin{tabular}{|c|c|c|}
\hline
Metrics: AUC & PTP  & RTP \tabularnewline
\hline
Interest & 0.723  &   0.706 \tabularnewline
\hline
Social  & 0.664 &  0.655 \tabularnewline
\hline
Demographic  & 0.713  &   0.691 \tabularnewline
\hline
Social + demographic & 0.76 &   0.759 \tabularnewline
\hline
Interest + demographic & 0.784  &  0.755 \tabularnewline
\hline
Interest + social & 0.766 &   0.75 \tabularnewline
\hline
Interest + social + demographic & \textbf{0.82} &   \textbf{0.81} \tabularnewline
\hline
\end{tabular}
\end{table}

\begin{table}[htb]
\centering
\caption{The performance of different feature combinations in the regression task}\label{tab:regression_diff_feature}
\begin{tabular}{|c|c|c|}
\hline
Metrics: reduced ratio of MAE & PTP  & RTP \tabularnewline
\hline
Interest & 14.5\%  &  11.29\% \tabularnewline
\hline
Social  & 10.25\% &  5.66\% \tabularnewline
\hline
Demographic  & 14.75\%  & 11.9\% \tabularnewline
\hline
Social + demographic & 18.48\% & 16.78\% \tabularnewline
\hline
Interest + demographic & 20.68\%  &  18.35\% \tabularnewline
\hline
Interest + social & 15.69\% &  13.07\% \tabularnewline
\hline
Interest + social + demographic & \textbf{23.88\%} &   \textbf{21.28\%} \tabularnewline
\hline
\end{tabular}
\end{table}

In general, each category of features would contribute to the interest similarity prediction on its own. And with more selected features, the prediction performance is better. Although various social relations are apparently correlated with interest similarity as show in Sec.~\ref{Sec:measurement}, they are not available to many user pairs. Moreover, the results show that for user pairs with partially available information, such as, only social and demographic information, we could still adopt the predictive models to improve the interest similarity estimation.

\section{Apply Our Findings to Recommendation}
\label{Sec:recommendation}
To demonstrate the practical value of the proposed prediction algorithms and the tag-based profiling scheme, we applied the predicted results to video recommendation. 
The goal of this section is to recommend a list of videos matching the user's video interests, that is, top-N recommendation~\cite{cremonesi2010performance}. 

For traditional recommendation algorithms, such as CF and content-based filtering, it is difficult to provide accurate recommendation for users with little or no recent interest information, which is known as the cold start problem~\cite{koren2009matrix}. However, if we could find some currently active users who are predicted to be similar to these inactive users, then we can recommend videos based on active users' interests to get over the cold start problem.
Similar to the principles of many existing solutions to address the cold start problem, we try to find closest neighbors (helpers) with similar interests to recommend items.

\subsection{Experiment Settings}
From our dataset, we randomly selected two thousand active users (as ground truth) on the target day, denoted by $\mathcal{U}^T$. 
To restrict the scope of neighbor selection without global searching from millions of users, we randomly drew five thousand candidate neighbors for each target user. To test the performance of social-friend-based approach, those candidate neighbors also include the friends of the target user.
For fair and unified comparison, we find top-K similar users from the candidate neighbors of each target user for recommending N videos by selecting the top-N popular videos among these K neighbors.

For neighbor selection in both tag-based (PTP \& RTP) and video-based profiling (VBP) schemes, we utilized the similarity values predicted by the regression tasks in Sec.~\ref{Sec:simPrediction}. 
Then the recommendation lists are generated by aggregating the selected neighbors' viewing records and keeping the top-N popular videos from those records. For comparison, we also implemented various algorithms corresponding to different strategies (both personalized and non-personalized) of closest neighbor selection. Three personalized approaches we used are as follows. 
\begin{itemize}
\item Demographic profile similarity: select K closest neighbors according to the similarity of demographic profiles.
\item Social friend filtering: select top-K close friends of the target user based on their interaction frequency, i.e., the monthly messaging days.
\item Past long-term profiling: when current interest information is unavailable, one conventional method is to utilize the past long-term records. In this method, we chose the top-K similar neighbors by comparing the past one month's video records of the target user and each candidate neighbor. 
\end{itemize}
Besides, we used two non-personalized algorithms as baselines.
\begin{itemize}
\item Random: randomly select K users from the candidate neighbors and recommend the most popular N videos from these users to each target user.
\item Global popularity: for each target user, always recommend the top-N popular videos among all the five thousand candidate neighbors.
\end{itemize}

\subsection{Evaluation Metrics}
To evaluate the performance of each algorithm, we employ two categories of metrics that measure not only the accuracy of recommendations, but also the diversity of the recommendation results. Assume the viewed video set of user $u$ is $\mathcal{I}_u$, and the video set recommended to user $u$ by a certain algorithm is $\mathcal{R}_u$.
\subsubsection{Accuracy.}
We utilize \textbf{F-measure}~\cite{powers2011evaluation} 
as the accuracy metrics which examines whether videos viewed by target users are ranked at top positions in the recommendation lists. 
\begin{itemize}
\item \textbf{F-measure} = $2*\frac{\mbox{Precisioon}*\mbox{Recall}}{\mbox{Precisioon} + \mbox{Recall}}$,
where $\mbox{Precision} =\frac{\sum_{u \in \mathcal{U}^T}|\mathcal{I}_u\cap \mathcal{R}_u| }{\sum_{u \in \mathcal{U}^T}|\mathcal{R}_u|} $; $\mbox{Recall} =\frac{\sum_{u \in \mathcal{U}^T}|\mathcal{I}_u\cap \mathcal{R}_u| }{\sum_{u \in \mathcal{U}^T}|\mathcal{I}_u|} $.
\end{itemize}

\subsubsection{Diversity.}
Many popular algorithms were reported to have a tendency to focus on popular items~\cite{jannach2013recommenders}, which is not helpful for discovering users' diverse interests. We use \textbf{Diversification} as the diversity metrics which measures the average inter-user difference of recommendation results~\cite{zhang2010personalized}. The larger this value, the more diverse the results are.
\begin{itemize}
\item $\textbf{Diversification} =1-\frac{2*\sum_{u,v\in \mathcal{U}^T}|\mathcal{R}_u \cap \mathcal{R}_v|}{|\mathcal{U}^T|*(|\mathcal{U}^T|-1)}$. 
\end{itemize}

\subsection{Experiment Results}
\subsubsection{Accuracy.}
\begin{figure}[htb]
  \centering
     \includegraphics[width=2.2in]{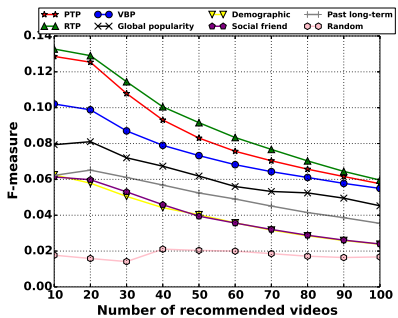}
	\caption{F-measure of each algorithm with different values of N when fixing K.}
  \label{Fig:accuracy_vs_N} 
\end{figure}

Since the number of videos viewed by each user varies, we tested the performance by different values of N (=$|\mathcal{R}_u|$). Typically, the length for the recommendation list is in tens. Thus, our experimental study focuses on the interval [10,100]. We firstly fixed the value of K (= 15), and varied N. The results\footnote{We obtained similar plots for other values of K.} are shown in Fig.~\ref{Fig:accuracy_vs_N}. The reason why the values of F-measure are relatively small is that we used historical data to conduct the experiments. Compared with online testing, there is an exposure bias for offline evaluation, that is, online users typically select and view videos from the list of videos exposed to them which are unknown to offline experiments. 

\begin{figure}[htb]
  \centering
     \includegraphics[width=2.2in]{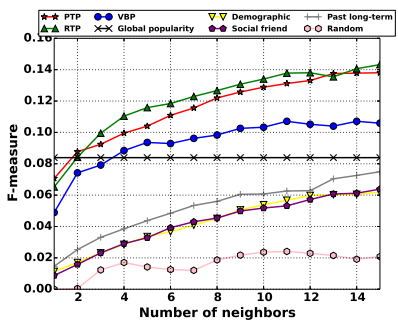}
	\caption{F-measure of each algorithm with different values of K when knowing the number of viewed videos.}
  \label{Fig:accuracy_vs_K} 
\end{figure}

Furthermore, one way to illustrate how different algorithms perform with regard to different number of neighbors, i.e, K, is to assume we know the number of videos viewed by each target user, which is equivalent to fixing N, and then vary the value of K. 
The corresponding results are in Fig.~\ref{Fig:accuracy_vs_K}. 

As shown in Fig.~\ref{Fig:accuracy_vs_N} and Fig.~\ref{Fig:accuracy_vs_K}, PTP and RTP could more accurately hit users' interests than VBP, which validated the advantage of the tag-based scheme over the video-based profiling scheme. Global popularity strategy is irrelevant to K since it consider all the candidate neighbors. Global popularity based recommendation could achieve moderate performance because it is derived from the statistics and popular videos are just what the majority of users have viewed. 
Actually, recent work also showed that popularity-based methods can represent a comparably strong baseline~\cite{steck2011item}. Moreover, the reason why the social friend based strategy did not perform well as expected is that there are fairly few friends as candidates compared to other cases. In other words, even interest similarity between friends is in general larger than that between random users, friends may not be the most similar users to the target users among all the users. In Fig.~\ref{Fig:accuracy_vs_K}, we found that when $K$ is small, such as one or two, algorthms did not perform well except for global popularity based approach. It's because, with a small number of neighbors, we might not have enough videos candidates for recommendation yet.

\subsubsection{Diversity.}
To further compare the diversity of recommendation results from tag-based and video-based profiling schemes, we tested the diversity metrics for different values of N and K. Due to space limitation, we only show two values of K, namely, 10 and 15. The results are shown in Table~\ref{tab:Diversification}.

\begin{table}[htb]
\centering
\caption{Diversification of each user profiling method with different values of K and N}\label{tab:Diversification}
\begin{tabular}{|c|c|c|c|c|c|c|}
\hline
\multirow{2}{*}{N=} &   \multicolumn{3}{|c|}{K=10}  &  \multicolumn{3}{|c|}{K=15} \\
\cline{2-7}
 & PTP & RTP & VBP  & PTP & RTP & VBP \\
\hline
 10 & 0.930	& 0.947	& 0.920	& 0.882	& 0.900	& 0.880 \\
\hline
 20 & 0.940 &	0.952 &	0.918&	0.903	&0.920	&0.871  \\
\hline
 30 &0.939 & 	0.951 & 	0.921 & 	0.917	 & 0.928	 & 0.878 \\
\hline
 40 &0.937 & 	0.950	 & 0.920	 & 0.920	 & 0.930 & 	0.883 \\
\hline
 50 &0.937 & 	0.950 & 	0.922 & 	0.925	 & 0.932 & 	0.887 \\
\hline
 60 &0.936	 & 0.947	 & 0.921 & 	0.925	 & 0.932	 & 0.890 \\
\hline
 70 &0.934	 & 0.946 & 	0.920	 & 0.925	 & 0.933 & 	0.892 \\
\hline
 80 &0.933 & 	0.944 & 	0.920 & 	0.925 & 	0.932 & 	0.893 \\
\hline
 90 &0.931	 & 0.943	 & 0.919	 & 0.926	 & 0.932	 & 0.894 \\
\hline
 100 & 0.930 & 	0.942	 & 0.918 & 	0.926 & 	0.932	 & 0.894 \\
\hline
\end{tabular}
\end{table}

Compared with recommendations from VBP, the results produced by two tag-based methods are more diverse in terms of inter-user difference.
Another conclusion is that, although, PTP is better than RTP with regard to prediction accuracy, RTP scheme could generate more diverse results, which is useful for discovery of the long-tail part of user interests. 

\section{Related Works}
\label{Sec:related}
With the popularity of OSNs, a better understanding of how much two individuals are alike in their interests,  namely, interest similarity, will benefit various applications in OSNs. For example, information about interest similarity can be leveraged to improve friend recommendation based on the observation that the like-minded users are more likely to become friends~\cite{lewis2012social}. Moreover, targeted online advertising can also benefit from this because we could largely expand the pool of potential clients by finding users who are similar to the existing clients~\cite{yu2014facebook}. In this paper, we apply the inferred interest similarity to video recommendation where the collaborative recommender systems identify users who are similar to a target user in order to recommend items to her.

There are many studies on user interest profiling. Users may express their interests explicitly in their profiles when they register online, such as facebook\cite{han2014alike}. Alternatively, user interest models can be derived by analyzing a user's own behaviors, which can be mainly divided into three categories: 1) item-based method which profiles users' interests by consumed items, such as viewed videos or collected books; 2) tag-based method where user interest profiles can be created by either aggregating the consumed item set and the tag sets of those items or directly taking the tags which the users have already used. \cite{hung2008tag} proposes a user profiling approach based on the tags associated with one's personal collection of contents. \cite{wang2011connecting} studies how to connect people share similar interests via a tag network. In collaborative tagging systems, such as del.icio.us\footnote{http://del.icio.us/}, users can choose their own words as tags to describe consumed items, which is also known as folksonomy~\cite{cai2010personalized}; 3) latent topic-based method which describes users' interests by some latent topics derived from matrix factorization~\cite{koren2009matrix} or LDA~\cite{krestel2009latent} from the video consumption and description data. In this paper, we adopt tag-based profiling which can alleviate the weakness of the item-based method and is more efficient in computation complexity for the large-scale practical system, i.e., Tencent Video.

When we do not have users' behavioral data, such as for new users, users' interests can only be inferred based other user information.
Recent studies start to leverage social cues to enhance user interest modeling. Trust is shown to be positively correlated with interest similarity~\cite{ziegler2007investigating}. Besides, people tend to befriend others who share similar traits, known as homophily in sociology~\cite{lewis2012social}. \cite{wen2010quality} deduces a user's interests by considering this user's social neighbors' interests. Also interests could be inferred to some extent from the users who share more demographic attributes~\cite{koren2009matrix}. A similar previous work investigated how various user information affects  interest similarity based on video-based user profiling~\cite{han2014alike}. In this work, we mine and learn to predict two users' interest similarity with two tag-based user profiling methods, namely, PTP and RTP, which are shown to be more reasonable and effective than the traditional video-based method. 

\section{Conclusion}
\label{Sec:conclusion}
In this paper, we mined and predicted users' interest similarity defined by tag-based interest profiles. We firstly systematically study the
correlation between interest similarity and various features, including demographic (age, gender and location), social (friendship, community membership, and interaction information) and interest (past long-term similarity and individuality of interest), to select the most effective features. Then we test the effectiveness of different feature combinations and machine learning models (including the hybrid tree-encoded linear model proposed by us) in predicting two users' interest similarity when their recent interests is few or blank. Furthermore, we apply our model to video recommendation to demonstrate its practical value. In the future work, we will try to explore and test more application scenarios for the interest similarity prediction. 

\selectfont
\bibliographystyle{aaai}
\bibliography{icwsm2016}
\end{document}